\documentclass{iopart}

\usepackage{graphics, rotating, iopams}
\usepackage[colorlinks]{hyperref}

\newtheorem{proposition}{Proposition}

\newtheorem{corollary}[proposition]{Corollary}

\def\d{\mathrm{d}}

\def \bn {\mbox{\boldmath{$n$}}}
\def \bk {\mbox{\boldmath{$k$}}}

\def \bl {\mbox{\boldmath{$\ell$}}}
\def \bm {\mbox{\boldmath{$m$}}}
\def \kS {{\cal H}}
\def\d{\mathrm{d}}

\newcommand{\pp}{{\it pp\,}-}

\newcommand{\be}{\begin{equation}}
\newcommand{\ee}{\end{equation}}
\newcommand{\bea}{\begin{eqnarray}}
\newcommand{\eea}{\end{eqnarray}}
\newcommand{\bean}{\begin{eqnarray*}}
\newcommand{\eean}{\end{eqnarray*}}

\def \D {\mathrm{D}}
\def \H {\mathcal{H}}
\newcommand{\M}[2] {{\stackrel{#1}{M}}_{#2}}

\newcommand{\half}{\frac{1}{2}}

\newcommand{\fracOmega}[1]{\frac{\Omega_{,#1}}{\Omega}}

\newcommand{\kronecker}[2]{\delta^{#1}_{#2}}

\newcommand{\bgmetric}[2]{\bar{g}_{#1 #2}}
\newcommand{\ibgmetric}[2]{\bar{g}^{#1 #2}}

\begin{document}
\title{Kerr--Schild spacetimes with (A)dS background}

\author{Tom\' a\v s M\'alek $^1$, Vojt\v ech Pravda  $^2$}
\address{$^1$ Institute of Theoretical Physics, Faculty of Mathematics and Physics,\\  Charles University,  
V Hole\v sovi\v ck\' ach 2, 180 00 Prague 8, Czech Republic}
\vspace{-1mm}
\eads{malek@math.cas.cz}

\vspace{3mm}
\address{$^2$ Institute of Mathematics, Academy of Sciences of the Czech Republic,\\  \v Zitn\' a 25, 115 67 Prague 1, Czech Republic}
\vspace{-1mm}
\eads{pravda@math.cas.cz}
\date{\today}


\begin{abstract}
General properties of Kerr--Schild spacetimes with (A)dS background in arbitrary dimension $n>3$ are studied. It is shown that  the geodetic Kerr--Schild vector $\bk$ is  a multiple WAND of the spacetime. Einstein Kerr--Schild spacetimes with non-expanding $\bk$ are shown to be of Weyl type N, while the expanding spacetimes are of type II or D. 

It is  shown that this class of spacetimes obeys the optical constraint. This allows us to solve Sachs equation, determine
$r$-dependence of boost weight zero components of the Weyl tensor and discuss curvature singularities.  
\end{abstract}


\section{Introduction}

Kerr--Schild (KS) class of spacetimes \cite{KerSch652}, i.e.\ metrics of the form
\be
g_{ab} = \eta_{ab} - 2 {\cal{H}} k_a k_b, 
\ee
with ${\cal{H}}$ being a scalar function and $\bk$ being a null vector with respect to the background flat metric $\eta_{ab}$ and full metric $g_{ab}$, play
an important role in the study of exact solutions of the vacuum Einstein equations in four and higher dimensions.
The  exceptional advantage of this ansatz  is that it makes analytic calculations tractable and allows analysis
of such spacetimes in full generality while at the same time it contains exact solutions of high interest, such as Kerr black holes and higher dimensional Myers--Perry black holes  \cite{MyePer86} and type N pp-waves \cite{Coleyetal06, OrtPraPra09}.
General properties of such metrics in arbitrary dimension were studied in \cite{OrtPraPra09}.

Rotating black holes with de Sitter and anti-de Sitter backgrounds discovered in four, five and higher dimensions in \cite{Carter68B}, \cite{Hawkingetal} and \cite{Gibbonsetal04_jgp}, respectively, can be cast to the generalized Kerr--Schild (GKS) form\footnote{See e.g.\ \cite{DerGur86,Gibbonsetal04_jgp} for the discussion of this class of metrics in higher dimensions.}
\be
g_{ab} = \bgmetric a b - 2 {\cal{H}} k_a k_b, \label{GKS}
\ee
with $\bk$ again being null vector with respect to background de Sitter or anti-de Sitter  metric $\bgmetric a b $ and full metric $g_{ab}$.

In this paper we analyze properties of metrics (\ref{GKS}) in dimension $n>3$ and generalize the main results
of \cite{OrtPraPra09} from the Ricci flat case to the case of Einstein spacetimes.
 Hereafter we thus assume that $\bgmetric a b = \Omega \eta_{ab}$ is $n$-dimensional (A)dS metric with  cosmological constant $\Lambda$,
with Minkowski metric $\eta_{ab}$ being in the canonical form $-{\rm d}t^2 + {{\rm d} x_1}^2 + \ldots + {{\rm d} x_{n-1}}^2$.

In section \ref{general_k} it is shown that under quite general conditions, including the case of Einstein spacetimes, Einstein equations
imply that the KS vector field $\bk$ is geodetic.
In section \ref{sec_curvature} curvature tensors and Einstein equations for the metric (\ref{GKS})
are studied in the case of geodetic $\bk$. It is also shown that $\bk$ is necessarily a multiple WAND.

In the rest of the paper we focus on Einstein GKS spacetimes. In section \ref{sec_Brinkmann}
we point out that Brinkmann warp product preserves GKS form.
In section \ref{sec_nonexp} it is  shown that  for the non-expanding $\bk$
Einstein GKS spacetimes belong to  type N Kundt class and explicit examples of such metrics are obtained using the Brinkmann warp product.
 In section \ref{sec_exp} we study the case with expanding  $\bk$.
It is shown that these spacetimes obey the ``optical constraint''  \cite{OrtPraPra09}. This allows us to determine $r$-dependence\footnote{With $r$ being
the affine parameter along KS congruence $\bk$.} of the optical matrix  and boost weight zero components of the curvature tensors and analyze
curvature singularities.

In section \ref{sec_disc} we briefly discuss the main results.
\ref{App_Weyl} contains frame components of Riemann and Weyl tensors in the case of geodetic KS vector $\bk$.   In the \ref{App_MP} we
compare the $r$-dependence of the optical matrix in a parallelly propagated frame  for general Einstein GKS metric and for  the five-dimensional (A)dS--Kerr black hole.

\subsection{Preliminaries}

Throughout the paper we  use standard notation of higher dimensional NP  formalism \cite{Pravdaetal04,OrtPraPra07} (see also \cite{GHPHD}).
For completeness, let us briefly summarize the notation and list several useful relations.

We will work in a real frame $\bn \equiv \bm^{(0)}$, $\bl \equiv \bm^{(1)}$, $\bm^{(i)}$
consisting of two null vectors $\bl$, $\bn$ and $n-2$ orthonormal spacelike vectors $\bm^{(i)}$ obeying
\be
\fl
\ell^a \ell_a = n^a n_a = \ell^a m_a^{(i)} = n^a m_a^{(i)} = 0 \, , \qquad
\ell^a n_a = 1 \, , \qquad
m^{(i)a} m_a^{(j)} = \delta_{ij} \, ,
\ee
with indices $i$, $j$, \ldots\ going from 2 to $n-1$ and $a$, $b$, \ldots\ from 0 to $n-1$.
Then the full metric takes the form
\be
g_{ab} = 2 n_{(a}\ell_{b)} + \delta_{ij} m_a^{(i)} m_b^{(j)} \, .
\ee
Throughout the paper we conveniently identify the KS vector $\bk$ with the null frame vector $\bl$. 

Ricci rotation coefficients $L_{ab}$, $N_{ab}$ and $\M {i} {bc}$ are defined as the frame components of covariant derivatives
\be
\fl
\ell_{a;b} = L_{cd} \, m_a^{(c)} m_b^{(d)} \, , \qquad
n_{a;b} = N_{cd} \, m_a^{(c)} m_b^{(d)} \, , \qquad
m_{a;b}^{(i)} = \M i {cd} \, m_a^{(c)} m_b^{(d)} \, .
\ee

In the case of geodetic and affinely parametrized vector $\bl$ the following definitions \cite{Pravdaetal04,OrtPraPra07}  are useful
\bea
& & S_{ij} \equiv L_{(ij)} =\sigma_{ij}+\theta\delta_{ij}, \qquad A_{ij}\equiv L_{[ij]} , \nonumber \label{optics} \\
& &  \theta\equiv\textstyle{\frac{1}{n-2}} S_{ii} , \qquad \sigma^2\equiv\sigma_{ij}\sigma_{ij}, \qquad \omega^2\equiv A_{ij}A_{ij},
\eea
where $S_{ij}$, $\sigma_{ij}$ and $A_{ij}$ are the {\em expansion}, {\em shear} and {\em twist} matrices, respectively, and  $\theta$, $\sigma$ and $\omega$ are the corresponding scalars.

Directional derivatives along the frame vectors are denoted as
\be
\D \equiv \ell^a \nabla_a \, , \quad
\Delta \equiv n^a \nabla_a \, , \quad
\delta_i \equiv m_{(i)}^a \nabla_a \, .
\ee

Finally,  the conformal factor $\Omega$ in the background de Sitter and anti-de Sitter  metric $\bgmetric a b = \Omega \eta_{ab}$ is
\bea
\Omega &=& \Omega^{+}=\frac{ {\ell^2_\Lambda}}{t^2} = \frac{(n-2)(n-1)}{2\Lambda t^2}, \\
\Omega &=& \Omega^{-} = \frac{a^2}{ {x_1}^2} = - \frac{(n-2)(n-1)}{2\Lambda {x_1}^2},
\eea
respectively, while
Minkowski limit $\Lambda = 0$ can be obtained by setting $\Omega = 1$.
Note also that  $\Omega$ satisfies
\be
\label{bgEFEs}
\frac{\Omega_{,ab}}{\Omega} = \frac{3}{2} \frac{\Omega_{,a} \Omega_{,b}}{\Omega^2}
\, , \qquad
- \frac{1}{4} \frac{\Omega_{,a} \Omega_{,b}}{\Omega^2} \ibgmetric a b = \frac{2}{(n-2)(n-1)} {\Lambda}.
\ee

When $\bk$ is geodetic and affinely parametrized, the following identities are also useful 
 \be
k^a_{\phantom{a};a} = L_{ii} \, , \qquad
k_{a;b} \, k^{a;b} = L_{ij} L_{ij} \, , \qquad
k_{a;b} \, k^{b;a} = L_{ij} L_{ji} \, .
\ee

\section{General KS vector field}
\label{general_k}

The main point of this section is to show that if energy--momentum tensor  obeys $T_{ab} k^a k^b=0$ then Einstein equations imply that KS vector 
field is geodetic. This fact is then used in the following sections. 

Inverse metric to (\ref{GKS}) has the form
\be
g^{ab} = \ibgmetric a b + 2 {\cal{H}} k^a k^b,     \label{iGKS}
\ee
where $\ibgmetric a b = \Omega^{-1} \eta^{ab}$.
Christoffel symbols read
\be
\fl
\Gamma^a_{bc} = - \left( \H k^a k_b \right)_{,c} - \left( \H k^a k_c \right)_{,b} + g^{as} \left( \H k_b k_c \right)_{,s} +  
\half \fracOmega c \kronecker a b + \half \fracOmega b \kronecker a c - \half \fracOmega s g^{a s} \bgmetric b c. \label{Christoffel}
\ee

When studying constraints following from the Einstein equations, 
 it is natural to start with the highest boost weight component of the Ricci tensor $R_{00} = R_{ab} k^a k^b$ --- since $\bk$ is present in
 $\Gamma^a_{bc}$,
 many terms in this contraction vanish. Though such calculation is  still quite involved it leads to a remarkably simple
 result
 \be
R_{00} = 2 \H k_{c;a} k^{a} k^{c}_{\phantom{c};b} k^{b} - \half (n-2) \left( \frac{\Omega_{,ab}}{\Omega} 
- \frac{3}{2} \frac{\Omega_{,a} \Omega_{,b}}{\Omega^2}  \right) k^{a} k^{b}
\ee
for general form of $\Omega$. Therefore for (A)dS background from (\ref{bgEFEs}) 
\be
R_{00} = 2 \H k_{c;a} k^{a} k^{c}_{\phantom{c};b} k^{b}.
\ee
From the Einstein equations  it now  follows
\begin{proposition}
\label{KSgeod}
 The null vector $k^a$ in the generalized Kerr--Schild metric (\ref{GKS}) is geodetic if and only if the component of the energy--momentum
tensor \mbox{$T_{00}={T_{ab} k^a k^b}$} vanishes.
\end{proposition}

Proposition \ref{KSgeod} implies that vector $\bk$ is geodetic  for Einstein GKS spacetimes.  In fact geodeticity of $\bk$  also holds for spacetimes with aligned matter fields such as aligned Maxwell field ($F_{ab} k^a \propto k_b$) or aligned pure radiation ($T_{ab} \propto k_a k_b$).  
Thus starting from section \ref{sec_curvature} we consider $\bk$ being {\em geodetic} and {\em affinely parametrized}.  This leads to a considerable simplification of the necessary calculations.

\subsection{KS congruence in the background  spacetime}

Here we point out that geodeticity and optical properties of the KS congruence in the background  (A)dS
spacetime and in the full GKS spacetime coincide.

Note that Christoffel symbols and curvature tensor components of the background (A)dS spacetime can be obtained
from the corresponding quantities in the full GKS spacetime  by simply setting ${\cal{H}}$ to zero.
Using (\ref{Christoffel}) it is straightforward to see that
\bea
k_{a;b} k^b &=& k_{a,b} k^b = k_{a\overline{;}b} k^b, \nonumber \\
k^a_{\phantom{a};b} k^b &=&  k^a_{\phantom{a},b} k^b + \fracOmega b k^a k^b = k^a_{\phantom{a}\overline{;}b} k^b,
\eea
where  $k_{a\overline{;}b}$ denotes a covariant derivative with respect to the background (A)dS metric $\bgmetric a b$.
Thus $\bk$ is geodetic in the full GKS metric iff it is  geodetic in the  (A)dS background $\bgmetric a b$. 

Following \cite{OrtPraPra09} we can introduce a null frame in the background $\bgmetric a b$ by replacing $\bn$ by $\tilde{\bn}$ and 
keeping remaining frame vectors unchanged
\be
 \tilde n_a=n_a+{\cal H} k_a, \label{background_n}
\ee
which guarantees
\be
 \bgmetric a b   = 2k_{(a} \tilde n_{b)}  + \delta_{ij} m^{(i)}_a m^{(j)}_b
\ee
and allows us to compare the optical matrices $L_{ij}$ and ${\tilde L_{ij}}$ in the full spacetime and in the background, respectively.
Note that for $\bk$  geodetic, $L_{ij}$ does not depend on our particular choice (\ref{background_n}) since
in such case $L_{ij}$ is invariant under null rotations with $\bk$ fixed \cite{OrtPraPra07}.

Using (\ref{Christoffel}) it follows
\be
 L_{ij}\equiv k_{a;b}m^{(i)a}m^{(j)b}=k_{a\overline{;} b} m^{(i)a}m^{(j)b} \equiv {\tilde L_{ij}}
\ee
and therefore the optical matrices of the congruence $\bk$ in the full GKS spacetime and in the (A)dS
background are equal.

\section{Curvature tensors for geodetic KS vector field}
\label{sec_curvature}

As discussed in section \ref{general_k}, for Einstein GKS spacetimes KS vector $\bk$ is always geodetic and therefore
from now on we assume geodeticity of $\bk$. Then we arrive at convenient expressions used in the following calculations
\be
\Gamma^a_{bc} k^b = - \D\H k^a k_c + \half \fracOmega c k^a + \half \fracOmega b k^b \kronecker a c - \half \fracOmega b \ibgmetric a b k_c,\ee

\be
\Gamma^a_{bc} k_a = \D\H k_b k_c + \half \fracOmega c k_b + \half \fracOmega b k_c - \half \fracOmega a k^a \bgmetric b c .
\ee

\subsection{Ricci tensor}
 
Ricci tensor of the GKS metric can be expressed as
\bea
  R_{ab} = \left( \H k_a k_b \right)_{;cd} g^{cd} - \left( \H k^s k_a \right)_{;bs} - \left( \H k^s k_b \right)_{;as}
+ \frac{2 {\Lambda}}{n-2} \bgmetric a b \nonumber \\
\qquad {}- 2 \H \left( \D^2 \H + L_{ii} \D\H + 2 \H \omega^2 \right) k_a k_b,  \label{Ricci}
\eea
 which for $\Lambda=0$ reduces to the  result of \cite{OrtPraPra09}.
 From (\ref{Ricci}) it follows that $\bk$ is an eigenvector of the Ricci tensor
\be
R_{ab} k^b = - \left[ D^2\kS + (n - 2)\theta D\kS + 2\kS\omega^2 - \frac{2 {\Lambda}}{n - 2} \right] k_a \label{eigenRicci}
\ee
 and  thus boost weight 1 frame components $R_{0i}$ of the Ricci tensor vanish along with $R_{00}$.
The non-vanishing frame components of the Ricci tensor read
\bea
\label{R01}
\fl
R_{01} = - \D^{2}\H - (n - 2) \theta \D\H - 2\, \H \omega^{2} + \frac{2 {\Lambda}}{n - 2}, \\
\label{Rij}
\fl
R_{ij} = 2 \H L_{ik} L_{jk} - 2 \left( \D\H + (n-2) \theta \H \right) S_{ij} + \frac{2 {\Lambda}}{n - 2} \delta_{ij}, \\
\fl
R_{1i} = - \delta_i (\D\H)
+ 2 L_{[i1]} \D\H
+ 2 L_{ij} \delta_j \H
- S_{jj} \delta_i \H \nonumber \\
\fl
\ \ 
{}+ 2 \H \left( \delta_j A_{ij}
+ A_{ij} \M{j}{kk}
- A_{jk} \M{i}{jk}
- L_{1i} S_{jj}
+ 3 L_{ij} L_{[1j]}
+ L_{ji} L_{(1j)} \right),  \label{R1i}\\
\fl
R_{11} = \delta_i (\delta_i\H)
+ \left( N_{ii} - 2 \H S_{ii} \right) \D\H
+ \left( 4 L_{1i} - 2 L_{i1} + \M{i}{jj} \right) \delta_i\H
- S_{ii} \Delta \H  \label{R11} + \frac{4 \H {\Lambda}}{n - 1} \nonumber \\
\fl
\ \
{}+ 2 \H \left( 2 \delta_i L_{[1i]} + 4 L_{1i} L_{[1i]} + L_{i1} L_{i1}
              - L_{11} S_{ii} + 2 L_{[1i]} \M{i}{jj} - 2 A_{ij} N_{ij} - 2 \H \omega^2 \right).
\eea

\subsection{Algebraic type of the Weyl tensor}

Components of the Weyl and Riemann tensor  for the GKS metric with geodetic $\bk$ are given in the \ref{App_Weyl}.
In the previous section we have seen  that positive boost weight frame components 
of the Ricci tensor identically vanish.
It turns out that this is true for the Weyl tensor as well, i.e.
\be
 C_{0i0j}=0 , \qquad C_{010i}=0, \qquad C_{0ijk}=0,
\ee
and therefore
\begin{proposition}
\label{GKS-Weyltypes}
Generalized Kerr--Schild spacetime (\ref{GKS}) with a geodetic Kerr--Schild vector $\bk$ is algebraically special with
$\bk$ being the multiple WAND. 
\end{proposition}

KS spacetimes (\ref{GKS}) with a geodetic Kerr--Schild vector $\bk$ are therefore of Weyl type II or more special.
Using also a result from \cite{PraPraOrt07} that spacetimes (not necessarily of the Kerr--Schild class) which are either static or stationary with non-vanishing ``expansion'' and ``reflection symmetry'' are compatible only with Weyl types G, I$_i$, D or O we immediately arrive at
\begin{corollary}
\label{GKS-Weyltypes2}
 Static generalized Kerr--Schild spacetimes (\ref{GKS}) with a geodetic Kerr--Schild vector $\bk$  are of type D or conformally flat. 
\end{corollary}
Similar statement also holds for the stationary case.
Note that the above proposition is not restricted to Einstein spaces --- the only assumption we need is that $\bk$ is geodetic, which is 
by proposition \ref{KSgeod} equivalent to $T_{00}={T_{ab} k^a k^b=0}$.

Note also that these results immediately imply that Kerr--de Sitter metrics in arbitrary dimension \cite{Gibbonsetal04_jgp} are of type D, as shown previously in \cite{Hamamotoetal06} by explicit calculation of the Weyl tensor.

\subsection{Vacuum Einstein equations}

Since all previous results were derived without imposing Einstein field equations we now proceed with
studying their implications for GKS spacetimes.  
From now on, we thus consider only Einstein spacetimes. 
Let us recall that in this case $\bk$ is necessarily geodetic by proposition \ref{KSgeod}. 
Vacuum Einstein field equations (with cosmological constant) read 
\be
\label{EFEs}
R_{ab} = \frac{2}{n-2} \Lambda g_{ab} .
\ee
Note that the terms containing cosmological constant ${\Lambda}$ in the boost weight zero Ricci components $R_{01}$ (\ref{R01}) and $R_{ij}$ (\ref{Rij}) cancel with the corresponding terms on the right hand side of the Einstein equations (\ref{EFEs}).
The frame components of Einstein vacuum equations thus read
\bea
\label{EFE01}
\D^{2}\H + (n - 2) \theta \D\H + 2\, \H \omega^{2} = 0, \\
\label{EFEij}
2 \H L_{ik} L_{jk} - 2 \left( \D\H + (n-2) \theta \H \right) S_{ij} = 0,\\
\label{EFE1i}
R_{1i} = 0 \, , \quad R_{11} = 0,
\eea
where $R_{1i}$ and $R_{11}$ are given by (\ref{R1i}) and (\ref{R11}), respectively.

Following \cite{OrtPraPra09}, we rewrite trace of (\ref{EFEij}) as
\bea
\label{EFEijTrace}
(n-2)\theta (\D \log\H) &= L_{ij} L_{ij} - (n-2)^2 \theta^2 \nonumber \\
&= \sigma^2 + \omega^2 - (n-2)(n-3) \theta^2.
\eea
Since $\H$ appears in (\ref{EFEijTrace}) only for $\theta \neq 0$ it is natural to study non-expanding KS solutions ($\theta=0$) 
and expanding KS solutions ($\theta \neq 0$) separately.
This will be done in sections \ref{sec_nonexp} and \ref{sec_exp}.

\section{Brinkmann warp product preserves GKS form}
\label{sec_Brinkmann}

In this section we point out that Brinkmann warp product (see \cite{Brinkmann, OPPwarp}) preserves GKS form. This warp product is thus a convenient
method for constructing $n$-dimensional Einstein GKS spacetimes from known  $(n-1)$-dimensional GKS Ricci-flat or 
Einstein metrics. This applies e.g.\ to black strings constructed from Myers--Perry black holes  with cosmological constant
or to Einstein Kundt metrics constructed from Ricci-flat/Einstein Kundt metrics (see \sref{sec_nonexp}).  

It follows from  \cite{Brinkmann}  that
starting with $(n-1)$-dimensional seed Einstein metric $\d\tilde s^2$, one can generate $n$-dimensional
Einstein metric $\d s^2$
\be
 \d s^2=\frac{1}{f(z)}\d z^2+f(z)\d\tilde s^2 ,  
 \label{ansatz} 
\ee
with
\be
 f(z)=-\lambda z^2+2dz+b , \qquad \lambda=\frac{2\Lambda}{(n-1)(n-2)} ,
 \label{warpfactor}
\ee
and with $b$ and $d$ being constant parameters. Necessary and sufficient condition for  $\d s^2$ being Einstein spacetime is that
\be
 \tilde R=(n-1)(n-2)(\lambda b+d^2),
\label{ricci-n-1}
\ee 
where $\tilde R$ is the Ricci scalar  of the  $(n-1)$-dimensional Einstein seed metric $\d\tilde s^2$. 
Note  that for $R\not=0$ only following combinations of signs of $\tilde R$ and $R$ are allowed: ($-$,$-$), (0,$-$), (+,$-$), (+,+) 
and that  only the case ($-$,$-$) is free from singularities at $f(z)=0$ 
  (see \cite{OPPwarp}) . 
 
It was also shown in \cite{OPPwarp} that Weyl type of  $\d s^2$ is the same or more special than the type
of $\d\tilde s^2$. In particular, if the seed metric $\d\tilde s^2$ is of type N then
$\d s^2$ is of the type N as well.

Furthermore, if the seed metric $\d\tilde s^2$ is of the GKS form (\ref{GKS}), then
$\d s^2$ given by 
\be
\frac{1}{f} \d z^2 + f  \bgmetric a b \d x^a \d x^b - 2 f {\cal{H}} k_a k_b \d x^a \d x^b.
\label{GKSwarped}
\ee
is also of the GKS form
since the new warped background metric ${f}^{-1} \d z^2 + f  \bgmetric a b \d x^a \d x^b$ is necessarily Einstein and conformally flat and therefore (A)dS or Minkowski. 

For setting the warped background metric to the canonical form, one may use following coordinate transformations:
\bea
\fl
\label{AdStoAdS}
\mathrm{AdS}_{n-1} \Rightarrow \mathrm{AdS}_{n}: \qquad
&x^2 = \tilde x^2 + \tilde z^2 , \qquad
&z = \frac{\sqrt{- \left( d^2 + \lambda b \right)} \tilde z}{\lambda \tilde x} + \frac{d}{\lambda} , \\
\fl
\mathrm{dS}_{n-1} \Rightarrow \mathrm{AdS}_{n}:
&t^2 = \tilde t^2 - \tilde z^2 ,\qquad
&z = \frac{\sqrt{d^2 + \lambda b} \, \tilde t}{- \lambda \tilde z} + \frac{d}{\lambda} , \\
\fl
\mathrm{dS}_{n-1} \Rightarrow \mathrm{dS}_{n}:
&t^2 = \tilde t^2 - \tilde z^2 , \qquad
&z = \frac{\sqrt{d^2 + \lambda b} \, \tilde z}{\lambda \tilde t} + \frac{d}{\lambda} , \\
\fl
\label{MtoAdS}
\mathrm{M}_{n-1} \Rightarrow \mathrm{AdS}_{n}:
&z = \frac{1}{\lambda \tilde z} + \frac{d}{\lambda} .
&
\eea

By an appropriate coordinate transformation one can set the warped metric (\ref{ansatz}) in a form conformal to a direct product \cite{OPPwarp}.  
Such form depends on the combination of signs of $\tilde R$ and $R$ and all such combinations are given in \cite{OPPwarp}.
 Here we list only cases relevant in this paper --- i.e.\ cases with  $R\not=0$ 
\bea
  \lambda>0: \qquad & \d s^2 =\cosh^{-2}(\sqrt{\lambda}x)(\d x^2+\d\tilde s^2) & (\tilde R>0) , \label{++b} \\
  \lambda<0: \qquad & \d s^2 =\cos^{-2}(\sqrt{-\lambda}x)(\d x^2+\d\tilde s^2) & (\tilde R<0) , \label{--b} \\
                    & \d s^2 =(-\lambda x^2)^{-1}(\d x^2+\d\tilde s^2) & (\tilde R=0) , \label{-0b} \\
                    & \d s^2 =\sinh^{-2}(\sqrt{-\lambda}x)(\d x^2+\d\tilde s^2) \qquad &(\tilde R>0) .	\label{-+b}	 									  									 
\eea
Note that $\tilde R$ and $\lambda$ are related by $|\tilde R|=(n-1)(n-2) |\lambda|$.

\section{Non-expanding GKS Einstein spacetimes}
\label{sec_nonexp}

Let us first consider Einstein GKS spacetimes with a non-expanding ($\theta = 0$) null KS congruence $\bk$.
From equation (\ref{EFEijTrace}) it follows that the congruence is also shear-free and twist-free $\sigma = \omega = 0$.
Thus in this case the optical matrix vanishes
\be
\label{nonexpand1}
L_{ij} = 0,
\ee
and Einstein equations (\ref{EFE01}) reduce to
\bea
\label{R01Kundt}
\D^{2}\H = 0, \\
\label{R1iKundt}
\delta_i (\D\H) - 2 L_{[i1]} \D\H = 0, \\
\delta_i (\delta_i\H) + N_{ii} \D\H
+ \left( 4 L_{1i} - 2 L_{i1} + \M{i}{jj} \right) \delta_i\H \nonumber \\
{}+ 2 \H \left( 2 \delta_i L_{[1i]} + 4 L_{1i} L_{[1i]} + L_{i1} L_{i1}
+ 2 L_{[1i]} \M{i}{jj}  \right) +   \frac{4 \H {\Lambda}}{n - 1}=0.
\label{R11Kundt}
\eea
From (\ref{nonexpand1})--(\ref{R1iKundt}) it follows that  all boost weight 0 and $-1$ Weyl components, as given in the \ref{App_Weyl}, vanish.
\begin{proposition}
\label{Prop_nonexp}
Einstein generalized Kerr--Schild spacetimes (\ref{GKS}) with non-expanding KS congruence $\bk$ are of type N with $\bk$ being the multiple WAND.
Twist and shear of the KS congruence $\bk$ necessarily vanish and these solutions thus belong to the class of Einstein type N Kundt spacetimes.
\end{proposition}

Note that the above statement is also valid if we admit an additional aligned null radiation term
in the Ricci tensor i.e.\ $R_{ab} = \frac{2 \Lambda}{n-2} g_{ab} + \Phi k_a k_b$.
The aligned null radiation term appears only on the right hand side of the Einstein field equation \eref{R11Kundt}
and therefore it does not affect the derivation of proposition \ref{Prop_nonexp}.

Kundt metrics defined as spacetimes admitting a null geodesic congruence with vanishing optical matrix $L_{ij}$ can be in $n$ dimensions expressed as
 \cite{Coleyetal06}
\be
\fl \d s^2 =2\d u\left[\d v+H(u,v,x^k)\d u+W_{i}(u,v,x^k)\d x^i\right]+ g_{ij}(u,x^k) \d x^i\d x^j , \label{Kundt_gen}
\ee
where $i,j=2 \dots n-1.$
In general Kundt spacetimes do not admit GKS form. This directly follows from the fact that there exist
e.g.\ type III Einstein Kundt spacetimes which, by proposition \ref{Prop_nonexp} are incompatible with GKS form.
It can be however shown~\cite{OrtPraPra09} that all type N Ricci-flat Kundt metrics \cite{Coleyetal06}
\be
 \fl \d s^2 =2\d u\left[\d v+H(u,v,x^k)\d u+W_{i}(u,v,x^k)\d x^i\right]+\delta_{ij}\d x^i\d x^j , \label{Kundt}
\ee
where functions $W_{i}$ and $H$ are given in \cite{Coleyetal06} admit KS form. Since there exist a  $H^0(r,u,x^k)$ for  which
metric (\ref{Kundt}) is flat, all metrics  (\ref{Kundt}) can be written in Kerr--Schild form ${\rm d}s^2={\rm d}s^2_{\mbox{\tiny flat}}+(H^{0}-H^{0}_{\mbox{\tiny flat}}){\rm d}u^2$ .

Let us now show, using results of  \cite{Ozsvathetal85,BiPo98} (see also \cite{Gripobook}), that all four-dimensional
type N Einstein Kundt spacetimes admit GKS form. This class of metrics can be expressed as  \cite{Ozsvathetal85} 
\begin{equation}
 \fl \d s^2 = -2 \frac{Q^2}{P^2} \d u \d v + \left( 2 k \frac{Q^2}{P^2} v^2  -  \frac{\left(Q^2\right)_{,u}}{P^2} v - \frac{Q}{P} H \right)  \d u^2
  + \frac{1}{P^2} \left( \d x^2 + \d y^2 \right)
  , \label{KN}
\end{equation}
where 
\bea 
P=1+\frac{\tilde \Lambda}{12} (x^2+y^2) , \quad
k=\frac{\tilde  \Lambda}{6} \alpha(u)^2+\frac{1}{2} \left(\beta(u)^2+\gamma(u)^2 \right) , \label{Kundt_k} \\
Q=\left(1-\frac{\tilde  \Lambda}{12} (x^2+y^2) \right) \alpha(u)+\beta(u) x+\gamma(u) y , \nonumber
\eea 
with $\tilde \Lambda$ being four-dimensional cosmological constant and $H=H(x,y,u)$.

These spacetimes are Einstein if
\be
P^2 (H_{,xx} + H_{,yy}) + \frac{2}{3}  \tilde \Lambda H = 0.  \label{KNEinstein}
\ee
The general solution of (\ref{KNEinstein}) is \cite{BiPo98}
\be
H = 2 f_{1,x} - \frac{\tilde  \Lambda}{3 P} (x f_1 + y f_2),
\ee
where functions $f_1=f_1(u,x,y)$ and $f_2=f_2(u,x,y)$ are subject to $f_{1,x}=f_{2,y}$,  $f_{1,y}=-f_{2,x}$.
It can be shown that metrics (\ref{KN}) are conformally flat for 
\be
H(x,y,u)=\frac{1}{P} \left(A \left(1-\frac{\tilde \Lambda}{12}   (x^2+y^2)\right)+B x + C y\right),
\ee
where $A(u)$, $B(u)$ and $C(u)$ are arbitrary functions. 
Thus all metrics (\ref{KN}) differ from the conformally flat case only by a factor of $\d u$  and are therefore GKS.

\subsection{Examples of higher dimensional GKS Einstein Kundt spacetimes} 

In this section we will use Brinkmann warp product discussed in \sref{sec_Brinkmann} to construct examples of higher dimensional Einstein Kundt spacetimes belonging to the GKS class. 

Let us first  use (\ref{-0b})  to construct $(n+1)$-dimensional  type N
generalized Kerr--Schild Einstein spacetimes   from  $n$-dimensional vacuum type N Kundt metrics (\ref{Kundt})
\be
\fl
\d s^2 = \frac{1}{-\lambda \tilde z^2} \left( 2\d u\left[\d v+H(u,v,x^k)\d u+W_{i}(u,v,x^k)\d x^i\right]+\delta_{ij}\d x^i\d x^j + \d\tilde z^2 \right) , \label{WarpedKundtConformal}
\ee
where $i,j=2 \dots n-1.$
By performing transformation $v = - \lambda \tilde v \tilde z^2$ we can set the above  metric to the canonical Kundt form (\ref{Kundt_gen})
\be
\fl
\d s^2 = 2 \d u \left[ \d\tilde v + \tilde H \d u
             + \tilde W_{\tilde \imath} \d x^{\tilde \imath} \right]
             + \frac{1}{-\lambda \tilde z^2} \delta_{\tilde \imath \tilde \jmath} \d x^{\tilde \imath} \d x^{\tilde \jmath} , \label{WarpedKundt}
\ee
where $\tilde \imath$, $\tilde \jmath=2 \dots n.$
\bea
\tilde H = \frac{1}{-\lambda \tilde z^2} H(u,v,x^k), \\
\tilde W_{ i} = \frac{1}{-\lambda \tilde z^2} W_{  i}(u,v,x^k) \, ,  \quad  i=1 \ldots n-2 \\
\tilde W_{(n-1)} = \frac{2 \tilde v}{\tilde z}, \\
\d x^{(n-1)} = \d\tilde z.
\eea

Vacuum type N Kundt spacetimes are VSI (all curvature invariants, including differential invariants constructed from arbitrary covariant derivatives of Riemann tensor vanish \cite{HDVSI}). In the case with non-vanishing cosmological  constant $\Lambda$ curvature invariants either vanish or are constants depending on $\Lambda$. All non-expanding Einstein Kerr--Schild spacetimes are thus CSI (metrics with constant scalar invariants) \cite{CSI}. In fact metrics (\ref{WarpedKundtConformal}), (\ref{WarpedKundt}) were already discussed in \cite{CSI,CSI_supergrav} in the context of CSI spacetimes and supergravity.  

So far we used only Ricci-flat type N seed metrics. One can however also warp Einstein seed metrics  (\ref{KN})  (Note that warping Einstein metrics (\ref{WarpedKundt}) does not lead to new results.) In principle one can use several possible combinations of signs of Ricci scalars of the seed metric and full metric (see \eref{++b}--\eref{-+b}) to construct a five-dimensional Einstein solutions from  (\ref{KN}). Note however that only the case \eref{--b} with both Ricci scalars being negative is free from curvature or parallelly propagated singularities at $f(z)=0$ \cite{OPPwarp}. Therefore here we limit ourselves to the seed metrics (\ref{KN}) with  $\tilde \Lambda<0$ and warp product \eref{--b}   which leads to  five dimensional metrics 
\bea
 \fl \d s^2 = \frac{1}{\cos^2 (\sqrt{-\frac{\tilde \Lambda}{3}} z)} \Biggl( -2 \frac{Q^2}{P^2} \d u \d v + \left( 2 k \frac{Q^2}{P^2} v^2  -  \frac{\left(Q^2\right)_{,u}}{P^2} v - \frac{Q}{P} H \right)  \d u^2 \nonumber \\
 {}+ \frac{1}{P^2} \left( \d x^2 + \d y^2 \right) + \d z^2 \Biggr).
   \label{KN5D}
\eea

The four-dimensional seed metrics --- type N Kundt spacetimes with $\tilde \Lambda<0$ can be split to three
geometrically distinct subclasses (see \cite{Gripobook}). Depending on whether $k$ is positive, negative or vanishing, we will denote these 
metrics as KN($\tilde \Lambda^{-}, k^+$), KN($\tilde \Lambda^{-}, k^-$) and KN($\tilde \Lambda^{-}, k^0$) (generalized Siklos waves), respectively.

KN($\tilde \Lambda^{-}, k^+$) spacetimes  with the canonical choice $\alpha=0$, $\beta=\sqrt{2}$, $\gamma=0$ are represented by the metric \eref{KN}
where the functions $Q$ and $k$ are given by
\be
Q = {\sqrt{2}} x , \qquad
k = 1 .
\ee
One may put the background AdS metric to the canonical form by performing the coordinate transformation 
\bea
  u& = \frac{Y \mp \sqrt{T^2 - X^2 - Z^2}}{a}, \qquad &T= \frac{a^2 \left( 2 - P \right)}{2xv}, \nonumber \\
  v& = \pm \frac{a}{2 \sqrt{T^2 - X^2 - Z^2}},        &X= \frac{a^2 P}{2xv}, \nonumber \\
  x& = \pm \frac{2 a \sqrt{T^2 - X^2 - Z^2}}{X + T},  &Y= \frac{a \left( 1 + 2uv \right)}{2v}, \nonumber \\
  y& = \frac{2 a Z}{X + T},                           &Z= \frac{ay}{2xv},
\eea
where $a = \sqrt{-\frac{3}{\tilde \Lambda}}$.

The case KN($\tilde \Lambda^{-}, k^-$)  is represented by canonical  choice $\alpha=1$, $\beta=0$, $\gamma=0$
leading to 
\be
Q = 1 - \frac{\tilde \Lambda}{12} (x^2 + y^2) , \qquad
k = \frac{\tilde \Lambda}{6} .
\ee
In this case
the background AdS metric can be cast to the canonical form by using coordinate transformation
\bea
  u& = \sqrt{2} \left( \pm \sqrt{X^2 + Y^2 + Z^2} - T \right), \qquad &T = \sqrt{2} \frac{a^2 - uv}{2v}, \nonumber \\
  v& = \pm \frac{a^2}{\sqrt{2} \sqrt{X^2 + Y^2 + Z^2}},               &X = \frac{\sqrt{2} a^2 P}{2Qv}, \nonumber \\
  x& = \frac{2 a Z}{X \pm \sqrt{X^2 + Y^2 + Z^2}},                    &Y = \frac{\sqrt{2}ax}{2Qv}, \nonumber \\
  y& = \frac{2 a Y}{X \pm \sqrt{X^2 + Y^2 + Z^2}},                    &Z = \frac{\sqrt{2}ay}{2Qv},
\eea
where $a = \sqrt{-\frac{3}{\tilde \Lambda}}$.

In the last case with seed metrics KN($\tilde \Lambda^{-}, k^0$)  the canonical choice is $\alpha=1$, $\beta = \sqrt{-\frac{1}{3} {\tilde \Lambda}} \cos\theta$ and
$\gamma = \sqrt{-\frac{1}{3} {\tilde \Lambda}} \sin\theta$. 

It is worth to note that in special case  when $\theta$
is independent of $u$ (Siklos waves)  one can obtain the same five-dimensional metric by either warping
appropriate Einstein four dimensional seed metric (\ref{KN}) using \eref{--b}  or by warping Ricci-flat {\pp~waves} using \eref{-0b}.
This is related to the fact that Siklos waves can be cast to a form conformal to pp-waves (see e.g.\ \cite{Gripobook} for details).

\section{Expanding Einstein spacetimes}
\label{sec_exp}

\subsection{Optical constraint}

As in \cite{OrtPraPra09},  for $\theta \neq 0$ one can express $\D \log\H$ from equation (\ref{EFEijTrace})
\be
\D \log\H = \frac{L_{ik} L_{ik}}{\theta (n-2)}  - (n-2) \theta, \label{DE_H}
\ee
 which after substituting back to (\ref{EFEij}) leads to the ``optical constraint'' \cite{OrtPraPra09}
\be
L_{ik} L_{jk} = \frac{L_{lk} L_{lk}}{(n-2) \theta} S_{ij}.
\ee
It follows that $L_{ij}$ is also  a normal matrix and thus it can be put into a block-diagonal form by appropriate spins.   
Furthermore, such canonical frame is compatible with parallel transport along  $\bk$ \cite{OrtPraPra10}. 
Consequently, dependence of the optical matrix on the affine parameter $r$ along $\bk$ can be determined
from Sachs equation \cite{OrtPraPra10,OrtPraPra09}. This leads to
\bea
\label{blockdiagL}
 L_{ij}=\left(\begin {array}{cccc} \fbox{${\cal L}_{(1)}$} & & &  \\
  & \ddots & & \\ 
 & & \fbox{${\cal L}_{(p)}$} & \label{L_general} \\
 & & & \fbox{$\begin {array}{ccc} & & \\ \ \ & \tilde{\cal L} \ \ & \\ & & \end {array}$}
  \end {array}
 \right) ,
\eea
with ${\cal L}_{(1)}, \dots ,{\cal L}_{(p)}$ being
$2\times 2$ blocks of the form
\bea
 & & {\cal L}_{(\mu)}=\left(\begin {array}{cc} s_{(2\mu)} & A_{2\mu,2\mu+1} \nonumber \\
 -A_{2\mu,2\mu+1} & s_{(2\mu)} 
\end {array}
 \right) \qquad (\mu=1,\ldots, p) , \\
  & & s_{(2\mu)}=\frac{r}{r^2+(a^0_{(2\mu)})^2} , \qquad A_{2\mu,2\mu+1}=\frac{a^0_{(2\mu)}}{r^2+(a^0_{(2\mu)})^2} , \label{s_A} 
\eea
and  $\tilde{\cal L}$ being  $(n-2-2p)\times(n-2-2p)$-dimensional diagonal matrix
\be
   \tilde{\cal L}=\frac{1}{r}\mbox{diag}(\underbrace{1,\ldots,1}_{(m-2p)},\underbrace{0,\ldots,0}_{(n-2-m)})
 \label{diagonal}
\ee
with $0\le 2p\le m\le n-2$ and $m$ denoting the rank of $L_{ij}$. 

As in \cite{OrtPraPra09} trace of $L_{ij}$ is
\be
  (n-2)\theta=2\sum_{\mu=1}^p\frac{r}{r^2+(a^0_{(2\mu)})^2}+\frac{m-2p}{r} \label{rdep-expansion}
\ee
and
\be
 L_{ik} L_{ik}=(n-2)\theta \frac{1}{r}.
\ee
Using the above results we can determine the $r$-dependence of ${\cal H}$ by integrating (\ref{DE_H})
\be
\label{rdepH}
 {\cal{H}}=\frac{{\cal{H}}_0}{ {r^{m-2p-1}}}\prod_{\mu=1}^p\frac{1}{r^2+(a^0_{(2\mu)})^2} ,
\ee
which is identical to the case with vanishing $\Lambda$ discussed in detail in \cite{OrtPraPra09}.

\subsection{Algebraic type}
Let us show that Weyl types III and N are not compatible with  expanding Einstein KS spacetimes.  

For types III and N, boost weight zero Weyl components vanish. In particular vanishing of $C_{0i1j}$ as given in 
\ref{App_Weyl} implies 
\be
  L_{ij} \D\H = 2 \H A_{ik} L_{kj}.
\ee
Multiplying the above equation with $L_{lj}$, using the optical constraint and taking the trace gives
\be
\theta \, \D\H =0. \label{expandingWeylcond}
\ee
Now we can repeat the argument given in Appendix B of \cite{OrtPraPra09} that case $\D\H =0$ implies
$A_{ij}=0$ and $S_{ij}={\mbox{diag}}(s,0,\dots,0)$. This form of the optical matrix is not compatible 
with the canonical form of $L_{ij}$ for Einstein spacetimes of types III and N determined in \cite{Pravdaetal04} using Bianchi identities
 in the vacuum case. 
Since  cosmological constant does not enter Bianchi identities, same results follow also for Einstein spacetimes.
Note that although in the corresponding proof  in \cite{Pravdaetal04}  additional assumptions were made in the type III
case, these assumptions were not used in the non-twisting case needed here. We can thus conclude that expanding Einstein GKS
solutions with  $\D\H =0$ do not exist. Then from (\ref{expandingWeylcond})

\begin{proposition}
\label{prop_expandingWeyltype}
Einstein generalized Kerr--Schild spacetimes (\ref{GKS}) with expanding KS congruence $\bk$ are of Weyl types II or D or conformally flat.
\end{proposition}

\subsection{r-dependence of b.w. 0 components}

For expressing $r$-dependence of boost weight zero components of the Weyl tensor
we adopt more compact notation  \cite{PraPraOrt07,GHPHD}, 
\be
\fl
\label{bw0Phi}
\Phi_{ij} \equiv C_{0i1j} \, , \qquad
\Phi = C_{0101} \, , \qquad
\Phi^S_{ij} = - \frac{1}{2} C_{ikjk} \, , \qquad
\Phi^A_{ij} = \frac{1}{2} C_{01ij} \, .
\ee
Substituting $r$-dependence of $L_{ij}$ (\ref{blockdiagL})--(\ref{diagonal}) to the expressions for  the corresponding
Weyl tensor components 
 given in \ref{App_Weyl} we immediately obtain $r$-dependence of $\Phi_{ij}$ 
\bea
\Phi_{2\mu, 2\mu} = \Phi_{2\mu + 1,2\mu + 1} = - \D\H s_{(2\mu)} - 2 \H A^2_{2\mu, 2\mu + 1} \, , \nonumber \\
\Phi_{2\mu, 2\mu + 1} = \Phi^A_{2\mu, 2\mu + 1} = - \D (\H A_{2\mu, 2\mu + 1})  \, , \label{rdepPhi} \\
\Phi_{\alpha \beta} = - r^{-1} \delta_{\alpha \beta} \, , \qquad \Phi = \D^2\H \, . \nonumber
\eea
Hence $\Phi_{ij}$ reproduces the block diagonal structure of matrix $L_{ij}$.
Similarly one can determine $r$-dependence of the remaining non-vanishing boost weight zero components
\bea
\fl
C_{2\mu, 2\mu+1, 2\mu, 2\mu+1} = 2 \H \left( 3 A^2_{2\mu, 2\mu+1} - s^2_{(2\mu)} \right) \, , \nonumber \\
\fl
C_{2\mu, 2\mu+1, 2\nu, 2\nu+1} = 2 C_{2\mu, 2\nu, 2\mu+1, 2\nu+1} = -2 C_{2\mu, 2\nu+1, 2\mu+1, 2\nu} =
4 \H A_{2\mu, 2\mu+1} A_{2\nu, 2\nu+1} \, , \nonumber \\
\fl
C_{2\mu, 2\nu, 2\mu, 2\nu} = C_{2\mu, 2\nu+1, 2\mu, 2\nu+1} = -2 \H s_{(2\mu)} s_{(2\nu)} \, , \nonumber \\
\fl
C_{(\alpha)(i)(\alpha)(i)} = - 2 \H s_{(i)} r^{-1} \, , \label{rdepCijkl}
\eea
where $\mu \neq \nu$.  

\subsection{Singularities}
\label{subsec_singularities}

Let us briefly discuss curvature singularities of Einstein expanding GKS metrics. Since these spacetimes
are by proposition \ref{prop_expandingWeyltype} of types II or D (omitting the trivial conformally flat case),
the Kretschmann scalar is determined by boost weight zero Weyl components 
\bea
\fl R_{abcd} R^{abcd} &=& 4 \left( R_{0101} \right)^2 - 4 R_{01ij} R_{01ij} + 8 R_{0i1j} R_{0j1i} + R_{ijkl} R_{ijkl} \\
\fl &=& 4 \Phi^2 + 8 \Phi^S_{ij} \Phi^S_{ij} - 24 \Phi^A_{ij} \Phi^A_{ij} + C_{ijkl} C_{ijkl} + \frac{8 n}{(n-1)(n-2)^2} \, \Lambda^2.
 \label{Kretschmann}
\eea
The only additional term with respect to the vacuum case is the last constant term proportional to $ \Lambda^2$, which clearly cannot influence
singularities of the expression. Therefore, using results of \cite{OrtPraPra09}, in the ``generic'' case ($2p \neq m$, $2p \neq m - 1$) 
curvature singularities are located at $r=0$. Note that this case also includes all expanding, non-twisting Einstein GKS solutions, such as 
higher dimensional (A)dS--Schwarzschild--Tangherlini black holes.

In the special cases $2p = m$ and  $2p = m - 1$, presence of curvature singularity depends on the behavior of  functions $a^0_{(2\mu)}$, which depend on other coordinates than $r$. If $a^0_{(2\mu)}$  admit real roots at $x=x_0$, then a curvature singularity is located at $r=0$, $x=x_0$. This case corresponds e.g.\ to the ring shaped singularity of the Kerr--de Sitter spacetime (see \ref{App_MP} for details).

\section{Summary and discussion}
\label{sec_disc}
Although corresponding calculations for Einstein GKS spacetimes are considerably more involved, 
most of the results originally obtained in the vacuum
case in \cite{OrtPraPra09} hold for non-vanishing cosmological constant as well. 

In particular the KS vector $k$ is geodetic iff $T_{00}=T_{ab} k^a k^b$ component of the stress energy tensor vanishes.
Since this holds for Einstein spacetimes, we have further assumed $\bk$ being geodetic.
It then can be shown that GKS spacetimes are algebraically special with $\bk$ being the multiple WAND. 

GKS metrics naturally split into two subclasses with expansion $\theta$ either vanishing or non-vanishing.

In the Ricci-flat case it has been shown that {\em non-expanding}
KS spacetimes are equivalent to the  Kundt type N solutions. It is not clear at present
whether such equivalence holds for Einstein GKS Kundt type N as well. Here we have just shown that such equivalence holds in four dimensions
and that in higher dimensions non-expanding Einstein GKS spacetimes belong
to Einstein Kundt type N.  We also constructed several explicit examples of  Einstein GKS Kundt spacetimes using the Brinkmann warp product.

It has been also shown that for {\em expanding} Einstein GKS spacetimes 
optical matrix $L_{ij}$ obeys the optical constraint. In combination with $\bk$ being a WAND, it allows us to solve Sachs equation 
(see \cite{OrtPraPra10} for related discussion in more general context), determine the $r$-dependence of the optical matrix (see \ref{App_MP} for
comparison of the general GKS case with the five-dimensional (A)dS--Kerr black hole), KS function $\H$,
boost weight zero components of the Weyl tensor and Kretschmann scalar. It has been also observed that in the non-twisting case 
a curvature singularity is always located at $r=0$ (this for example applies to higher dimensional (A)dS--Schwarzschild--Tangherlini black holes), while in some  twisting cases further information is needed (note that e.g.\ five-dimensional Kerr--de Sitter black hole with two non-zero spins
is regular at $r=0$, while it is singular when one spin vanishes, see \ref{App_MP} for details).

In future works it would be of interest to study whether some of the above results hold in more general context, such as
for Kerr--Schild spacetimes in Einstein--Gauss--Bonnet gravity \cite{Anabalonetal}, for extended Kerr--Schild ansatz \cite{AlievCiftci09} (see also \cite{EttKastor2010}) or for multi-Kerr--Schild form \cite{CheLu08} and analyze
 what precisely are the conditions for these classes of spacetimes to admit some sort of hidden symmetries \cite{Krtousetal2008}.
 
It would be also useful to employ the results of this paper for finding new expanding Einstein GKS solutions or studying possible uniqueness of 
higher dimensional  (A)dS--Kerr black holes and related black strings/branes within this class of spacetimes.

\ack 
Some calculations in this paper were performed and many others checked using computer algebra software
Cadabra \cite{PeetersA,PeetersB}.
This work has been supported by research plan No AV0Z10190503. V.P. is supported by research grant
GA\v CR P203/10/0749, T. M. is supported by the project SVV 261301 of the Charles University in Prague.


\appendix
\section{Riemann and Weyl components}
\label{App_Weyl}
Riemann tensor frame components sorted by boost weight for geodetic and affinely parametrized KS vector $\bk$ read
\bea
\fl
R_{0i0j} = 0 \, , \qquad
R_{010i} = 0 \, , \qquad
R_{0ijk} = 0 \, , \label{R0i0j} \\
\fl
R_{0101} = \D^2\H - \frac{2 \Lambda}{(n - 2)(n - 1)} \, , \qquad
R_{01ij} = - 2 A_{ij} \D\H + 4 \H S_{k[j} A_{i]k} \, , \label{R01ij} \\
\fl
R_{0i1j} = - L_{ij} \D\H + 2 \H A_{ik} L_{kj}
           + \frac{2 \Lambda}{(n - 2)(n - 1)} \delta_{ij} \, , \\
\fl
R_{ijkl} = 4 \H \left( A_{ij} A_{kl} + A_{l[i} A_{j]k} + S_{l[i} S_{j]k} \right) 
+ \frac{2 \Lambda}{(n - 2)(n - 1)} \left( \delta_{ik} \delta_{jl} - \delta_{il} \delta_{jk} \right) \, , \label{Rijkl} \\
\fl
R_{011i} = - \delta_i \left( \D\H \right) + 2 L_{[i1]} \D\H + L_{ji} \delta_j\H + 2 \H \left( L_{1j} L_{ji} - L_{j1} S_{ij} \right) \, , \\
\fl
R_{1ijk} = 2 L_{[j|i} \delta_{|k]}\H + 2 A_{jk} \delta_i\H
           + 4 \H \Big( \delta_{[k} S_{j]i} + \M{l}{[jk]} S_{il} - \M{l}{i[j} S_{k]l} \nonumber \\
	   {}+ L_{1i} A_{jk} + L_{1[k} A_{j]i} \Big) \, , \\
\fl
R_{1i1j} = \delta_i ( \delta_j\H )
           + \M{k}{(ij)} \delta_k\H
           + 4 L_{1(i} \delta_{j)}\H  - 2 L_{(i|1} \delta_{j)}\H 
           + N_{(ij)} \D\H
           - S_{ij} \Delta\H \nonumber \\
	   {}+ 2 \H \Big( \delta_{(i} L_{1|j)}
                         - \Delta S_{ij}
                         - 2 L_{1(i} L_{j)1}
			 + 2 L_{1i} L_{1j}
                         - L_{k(i} N_{k|j)} 
                         + L_{1k} \M{k}{(ij)} \nonumber \\
           {}- 2 \H L_{k(i} A_{j)k}
			 - 2 \H A_{ik} A_{jk}
			 - L_{k(i} \M{k}{j)1} 
			 - L_{(i|k} \M{k}{j)1} \Big) \, .
\eea

Weyl frame components for GKS Einstein spaces (\ref{EFEs}) 
are 
\bea
C_{0i0j} = 0 \, , \qquad
  C_{010i} = 0 \, , \qquad
  C_{0ijk} = 0 \, , \\
C_{0101} = R_{0101} + \frac{2 \Lambda}{(n - 2)(n - 1)} \, , \qquad \label{C0101}
  C_{01ij} = R_{01ij} \, , \\
C_{0i1j} = R_{0i1j} - \frac{2 \Lambda}{(n - 2)(n - 1)} \delta_{ij} \, , \\
C_{ijkl} = R_{ijkl} - \frac{2 \Lambda}{(n - 2)(n - 1)} \left( \delta_{ik} \delta_{jl} - \delta_{il} \delta_{jk} \right) \label{C0i1j} \, , \label{Cijkl} \\
C_{011i} = R_{011i} \, , \qquad
  C_{1ijk} = R_{1ijk} \, , \qquad
  C_{1i1j} = R_{1i1j} \, . \label{C1i1j}
\eea

\section{Five-dimensional Kerr--(A)dS metric}
\label{App_MP}

Higher dimensional Kerr--(A)dS metric  in the GKS form \eref{GKS} is given in \cite{Gibbonsetal04_jgp}.
 In five dimensions the background metric, KS vector $\bk$
and function $\mathcal{H}$ are
\bea
\fl
  \bar{\mathbf g} = -\frac{(1-\lambda r^2) \Delta}{(1 + \lambda a^2)(1 + \lambda b^2)} \, \d t^2
    +\frac{r^2 \rho^2}{(1 - \lambda r^2)(r^2 + a^2)(r^2 + b^2)} \, \d r^2
    +\frac{\rho^2}{\Delta} \, \d \theta^2 \nonumber \\ 
    {}+ \frac{(r^2 + a^2) \sin^2\theta}{1 + \lambda a^2} \, \d \phi^2
    +\frac{(r^2 + b^2) \cos^2\theta}{1 + \lambda b^2} \, \ \d \psi^2 \, , \nonumber \\
\fl
  \bk =  \frac{\Delta}{(1 + \lambda a^2)(1 + \lambda b^2)} \, \d t
    + \frac{r^2 \rho^2}{(1 - \lambda r^2)(r^2 + a^2)(r^2 + b^2)} \, \d r \nonumber \\
    {}- \frac{a \sin^2\theta}{1 + \lambda a^2} \, \d \phi
    - \frac{b \cos^2\theta}{1 + \lambda b^2} \, \d \psi \, , \nonumber \\
\fl
  \mathcal{H} = - \frac{M}{\rho^2} \, ,
\eea
where
\bea
  \rho^2 = r^2 + \nu^2 \, , \quad
  \Delta = 1 + \lambda \nu^2 \, , \quad \nu = \sqrt{a^2 \cos^2\theta + b^2 \sin^2\theta} \,
\eea
and $\lambda$ is defined as in \eref{warpfactor}.

In agreement with propositions \ref{KSgeod}
and \ref{GKS-Weyltypes}, the KS vector $\bk$ is a geodetic multiple WAND. In fact  $\bk$ is also   affinely parametrized.
Let us complete a null frame by choosing the following  null vector $\bn$ and spacelike vectors $\bm^{(i)}$  
\bea
\fl
  \bk = - \frac{1}{1 - \lambda r^2} \frac{\partial}{\partial t}
       + \frac{\partial}{\partial r}
       - \frac{a}{r^2 + a^2} \frac{\partial}{\partial \varphi}
       - \frac{b}{r^2 + b^2} \frac{\partial}{\partial \psi} \ , \nonumber \\
\fl
  \bn = \left( \frac{1}{2} \frac{(1 + \lambda a^2)(1 + \lambda b^2)(1 - \lambda r^2)}{\Delta}
       - \frac{M}{\rho^2} \right) \bk
       + \frac{(1 + \lambda a^2)(1 + \lambda b^2)}{\Delta} \frac{\partial}{\partial t} \, , \nonumber \\
\fl
  \bm^{(2)} = \frac{\sqrt{\Delta}}{\rho} \frac{\partial}{\partial \theta} \, , \nonumber \\
\fl
  \bm^{(3)} = \frac{\rho \sin \theta \cos \theta}{\sqrt{\Delta} \nu }
       \bigg[ \frac{(b^2 - a^2)(1 - \lambda r^2)}{\rho^2} \frac{\partial}{\partial r}
       - \frac{a(1 + \lambda a^2)}{(r^2 + a^2) \sin^2 \theta} \frac{\partial}{\partial \varphi} \nonumber \\
       {}+ \frac{b(1 + \lambda b^2)}{(r^2 + b^2) \cos^2 \theta} \frac{\partial}{\partial \psi} \bigg], \nonumber \\
\fl
  \bm^{(4)} = \frac{a b r}{\nu}
       \left[ \frac{1 - \lambda r^2}{r^2} \frac{\partial}{\partial r}
       + \frac{1 + \lambda a^2}{a (r^2 + a^2)} \frac{\partial}{\partial \varphi}
       + \frac{1 + \lambda b^2}{b (r^2 + b^2)} \frac{\partial}{\partial \psi} \right] \, ,
  \label{KerrAdS-frame}
\eea
such that the optical matrix $L_{ij}$ takes the block-diagonal form \eref{blockdiagL}. 
We then  find a parallelly propagated frame by transforming the frame \eref{KerrAdS-frame} and  requiring 
that the block-diagonal structure of $L_{ij}$ remains unchanged. This can be achieved by  performing a rotation in $\bm^{(2)}$, $\bm^{(3)}$ plane
followed by a null rotation with fixed $\bk$
\bea
  \hat{\bn} = \bn + z_2 \hat{\bm}^{(2)} + z_4 \hat{\bm}^{(4)} 
    + \frac{1}{2} \left( z_2^2 + z_4^2 \right) \bk , \nonumber \\
  \hat{\bm}^{(2)} =\frac{\nu}{\rho} \bm^{(2)} - \frac{r}{\rho} \bm^{(3)} - z_2 \bk , \nonumber \\
  \hat{\bm}^{(3)} = \frac{r}{\rho} \bm^{(2)} + \frac{\nu}{\rho} \bm^{(3)}  , \nonumber \\
  \hat{\bm}^{(4)} = \bm^{(4)} - z_4 \bk ,
  \label{KerrAdS-pp-frame}
\eea
with
\be
  z_2 = - \frac{\lambda (a^2 - b^2) r \sin \theta \cos \theta}{\nu} \, , \quad
  z_4 = - \frac{\lambda a b r}{\nu} \, .
\ee

The optical matrix $L_{ij}$ of five-dimensional Kerr--(A)dS metric is of rank $m=3$
and it contains one 2 $\times$ 2 block ($p=1$).
One may compare this particular $L_{ij}$ with the corresponding optical
matrix \eref{L_general} of general GKS spacetime (with $n=5$, $m=3$, $p=1$), see \tref{tab:comparison}.
The two presented quantities are in agreement and obviously
\be
  a^0_{(2)} = \nu \, , \qquad
  \mathcal{H}_0 = - M \, .
\ee

Let us  briefly discuss presence
of curvature singularities using the results of \sref{subsec_singularities}. 
If  $a \neq 0$, $b \neq 0$ then  $2p=m-1$ and since  $a^0_{(2)}$ does not admit roots,  there are no curvature singularities in this case.
If we set one of the spins to zero, e.g.\ $b = 0$, then $a^0_{(2)}$ has one real root at
$\theta = \frac{\pi}{2}$ corresponding to a ring shaped singularity
known from the four-dimensional Kerr solution.

Putting $a = b = 0$ (non-twisting case corresponding to (A)dS--Schwarzschild--Tangherlini limit) 
implies $p=0$. Since neither $2p=m-1$ nor $2p = m$ a curvature
 singularity is located at $r=0$.

\begin{table}
  \centering
  \begin{tabular}{c|c}
     5D Kerr--de Sitter & GKS ($n = 5$, $m = 3$, $p = 1$) \\
     \hline & \\
     $L = \left( \begin{array}{ccc} \frac{r}{\rho^2} & \frac{\nu}{\rho^2} & \phantom{a}0\phantom{a} \\[0.6em]
       -\frac{\nu}{\rho^2} & \frac{r}{\rho^2} & 0 \\[0.6em]
       0 & 0 & \frac{1}{r}
     \end{array} \right)$ &
     $L = \left( \begin{array}{ccc} s_{(2)} & A_{2,3} & \phantom{ab}0\phantom{ab} \\[0.9em]
       - A_{2,3} & s_{(2)} & 0 \\[0.9em]
       0 & 0 & \frac{1}{r}
     \end{array} \right) $ \\
     & \\
     $\mathcal{H} = - M \frac{1}{r^2 + \nu^2}$
     & $\mathcal{H} = \mathcal{H}_0 \frac{1}{r^2 + (a^0_{(2)})^2}$ \\[1em]
     $s_{(2)}= \frac{r}{r^2 + \nu^2}$
     & $s_{(2)} = \frac{r}{r^2 + (a^0_{(2)})^2}$ \\[1em]
     $A_{2, 3} = \frac{\nu}{r^2 + \nu^2}$
     & $A_{2, 3} = \frac{a^0_{(2)}}{r^2 + (a^0_{(2)})^2}$
  \end{tabular}
  \caption{Comparison of the optical matrices of five-dimensional Kerr--(Anti-)de Sitter
  and corresponding generalized Kerr--Schild spacetime.}
  \label{tab:comparison}
\end{table}

\end{document}